\documentclass{aa}
\usepackage{graphicx,psfrag,amssymb,amsmath}
\usepackage{txfonts}
\usepackage{natbib}
\usepackage{epsfig}
\bibpunct{(}{)}{;}{a}{}{,}
\begin{document}
\title{Line profiles of water for the photon dominated region and embedded sources in the S140 region}
\subtitle{}
\author{Dieter R.~Poelman
           \inst{1,2}
           \and
        Marco Spaans
           \inst{1}
        }
\offprints{D.R.Poelman@astro.rug.nl}
\institute{Kapteyn Astronomical Institute, P.O. Box 800, 9700 AV Groningen, the Netherlands\\
\email{D.R.Poelman@astro.rug.nl}
        \and
         SRON Netherlands Institute for Space Research, Landleven 12, 9747 AD Groningen, the Netherlands
           }
\date{Received / Accepted}
\abstract
{$\mathrm{H_2O}$ is a key ingredient in many interstellar environments, like photon dominated regions and star forming clouds. It plays an important role in the oxygen chemistry and can act as a coolant in dense interstellar clouds and shocks. Observations and modelling of water lines thus provide powerful diagnostics of the physical conditions in interstellar emission zones. A radiative transfer method for the treatment of molecular lines is presented. We apply this method to previous SWAS and ISO observations of water vapor in the source S140 in order to make models to plan for, and to interpret, HIFI data. Level populations are calculated with the use of a three-dimensional (multi-zone) escape probability method and with a long characteristic code that uses Monte Carlo techniques with fixed directions. Homogeneous and inhomogeneous models are used to compute the differences between $\mathrm{H_2O}$ line profiles across the S140 region. We find that when an outflow or infall velocity field with a gradient of a few \mbox{$\mathrm{km}\ \mathrm{s}^{-1}$} is adopted, line profiles with a FWHM of 6 \mbox{$\mathrm{km}\ \mathrm{s}^{-1}$} are found, in agreement with observations. Inhomogeneous models are favoured to produce a single-peaked line profile. When zooming in on smaller regions within the PDR, the shapes of the line profiles start to differ due to the different temperature and density distributions there. The embedded sources are traced by high excitation lines of, e.g., $3_{21}-2_{21}$, $3_{03}-2_{12}$, $2_{12}-1_{01}$ and $2_{20}-1_{11}$. The computed intensities are roughly consistent with existing ISO observations. Water emission in a PDR source like S140 requires a combination of a pure PDR and an embedded source in order to match the observations. Because of its good angular resolution, HIFI will be able to distinguish between a dense star forming region or a more diffuse gas component. It is therefore important for future observing programs to consider both in their predictions of the emission characteristics of water in these environments.
}
\titlerunning{Line profiles of water in the S140 region}
\maketitle
\section{\label{sec:intro}Introduction}
Our understanding of astrophysical objects, such as molecular clouds, young stellar objects (YSOs), disks, planetary nebulae and photon dominated regions (PDRs), requires both observations and, in order to retrieve information on physical parameters, detailed modelling. Line emission from molecules, such as \element[][]{CO} and \element[][]{H_2O}, has been observed extensively in the past decades \citep{1980ApJ...242..545F, 1996A&A...315L.173H, 2000ApJ...530L.129H, 1998ApJ...502..315H, 1998ApJ...495..871L, 2004ApJ...608..341S, 2004ARA&A..42..119V} and has been proven to be a powerful tool to reveal the behaviour of the aformentioned environments. Line profiles of molecules contain important information on the conditions of the regions they occupy, such as the density, and the velocity structure. Retrieving information from these lines requires accurate modelling. However, the interpretation of the line shapes has to be treated with great caution. \cite{2001MNRAS.326.1423R} noted that the classic signature of infall (an asymmetric double-peaked line which possesses a blue wing that is stronger than the red wing) could dissapear when molecules are depleted in spite of infall velocity fields. Moreover, \cite{2004ApJ...617..360L} argued that line observations can be misunderstood if one does not consider the variations in abundance with time or with radius. New instruments are needed to zoom in on these regions and spatially resolve them in order to retrieve more accurate information on their physical and chemical conditions. The launch of  Herschel, with on board the {\em Heterodyne Instrument for the Far Infrared} (HIFI), will make it possible to observe the Far Infrared world in more detail than ever before with higher angular resolution and sensitivity. As more infrared and submillimeter surveys are planned in the future (e.g., ALMA), more models to interpret the data from these instruments are needed.\\
One way to model the line emission from various molecules is to make use of an escape probability method, in which the probability $\beta$ indicates the chance for a photon to escape the cloud. This method, in which one solves locally the equations of statistical equilibrium in every gridpoint, has been used in the past to calculate, e.g., the radiative cooling of warm molecular gas \citep{1993ApJ...418..263N}, the far-infrared water emission from shock waves \citep{1996ApJ...456..611K} and water maser emission \citep{1991ApJ...368..215N}. Despite the wide range of applications of this method, other techniques, e.g., Monte Carlo (MC), Accelerated Lambda Iteration (ALI) and Local linearization (MULTI type) codes, that solve non-LTE multi-level radiative transfer problems, were created in order to meet with the non-local effects that arise in certain problems. Monte Carlo codes have been used over the years to study, e.g., masers \citep{1992MNRAS.258..159S}, UV continuum transfer \citep{1996A&A...307..271S} and molecular line transfer \citep{2000A&A...362..697H}. These have been proven to be reliable tools. However they have the disadvantage of having convergence problems at high optical depths ($\tau$ $\ge$ 100). \\
Recently, \cite{2006MNRAS.365..779E} introduced a new method to treat the radiative transfer of lines. This method, the Coupled Escape Probability (CEP), combines the algebraic equations used in the escape probability approximation with a closed set of equations in which the effect on the level populations in some zone {\em i} of radiation produced in all other zones is taken into account. Hence, the CEP method retains all the advantages of the escape probability approach. An eloquent discussion can be found in \cite{2006MNRAS.365..779E}. Our method, developed independently of the CEP, resembles the latter in that we also make use of a multi-zone formalism. However, the non-local component in the CEP treatment, i.e., the second term of equation (35) in \cite{2006MNRAS.365..779E}, is replaced in our formalism by a spatial iteration loop on the different escape probabilities in every zone along a number of discrete directions \citep{2005A&A...440..559P}. Note as well that our formalism already is fully 3D.\\
In this work, the multi-zone escape probability method of \cite{2005A&A...440..559P} is used to compute, through a ray-trace technique, line profiles for the \object{S140} region. The models of \cite{2005A&A...440..559P} for the S140 PDR are extended through the inclusion of embedded sources. For details on the physical structure of the S140 PDR and the employed escape probability formalism, we refer to \cite{2005A&A...440..559P}.\\
In Section \ref{sec:embsou} we discuss the excitation of certain water lines in an embedded source. Section \ref{sec:S140} describes the line profiles in S140 for the ground state transition of ortho-water. Also, in the appendix we test our escape probability model against a benchmark line profile problem and present a long characteristic radiative transfer method.
\section{\label{sec:embsou}Embedded Sources}
\begin{figure}[h!]
\begin{minipage}[c]{0.5\textwidth}
\includegraphics[width=4.5cm]{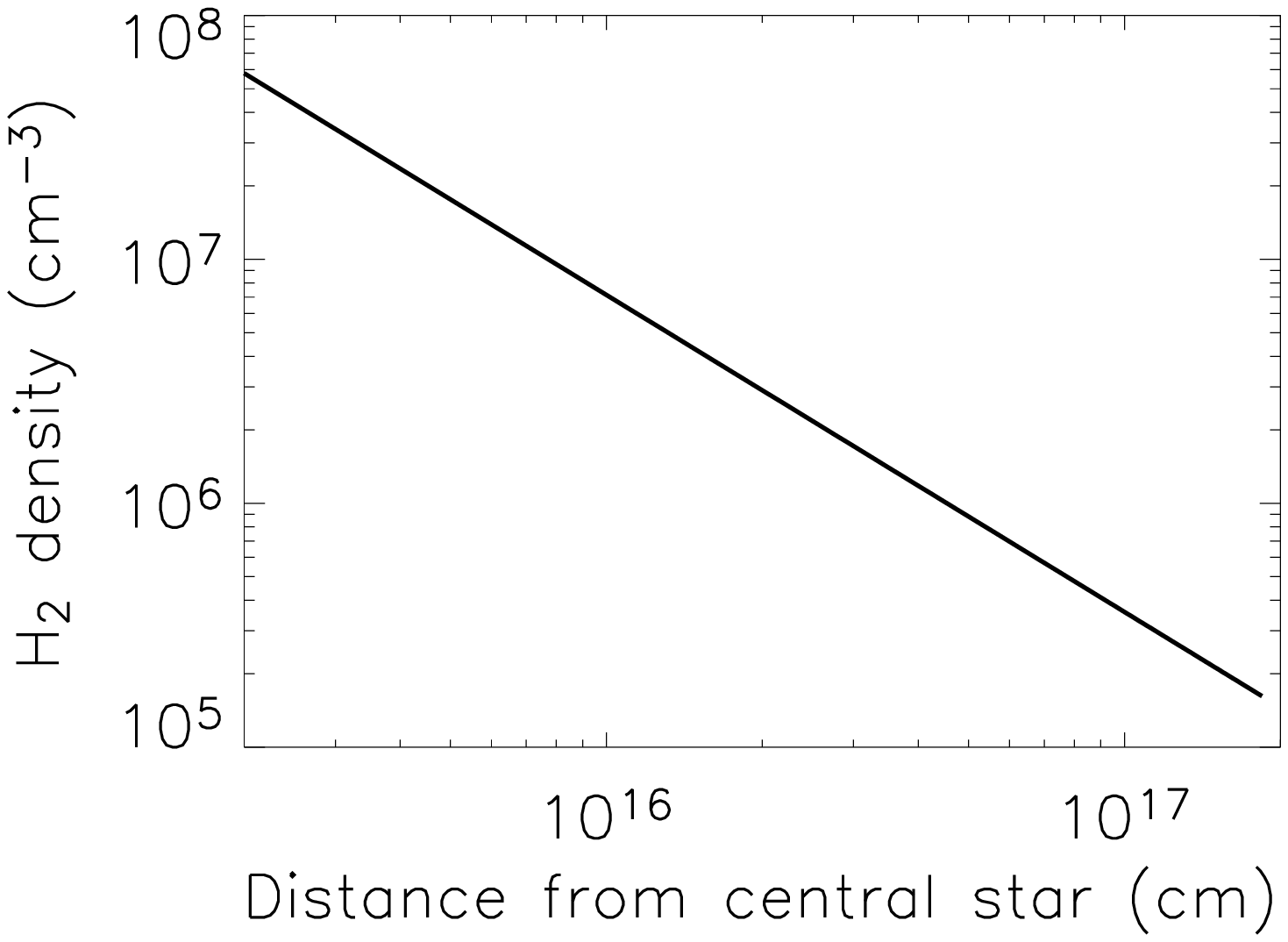}
\includegraphics[width=4.5cm]{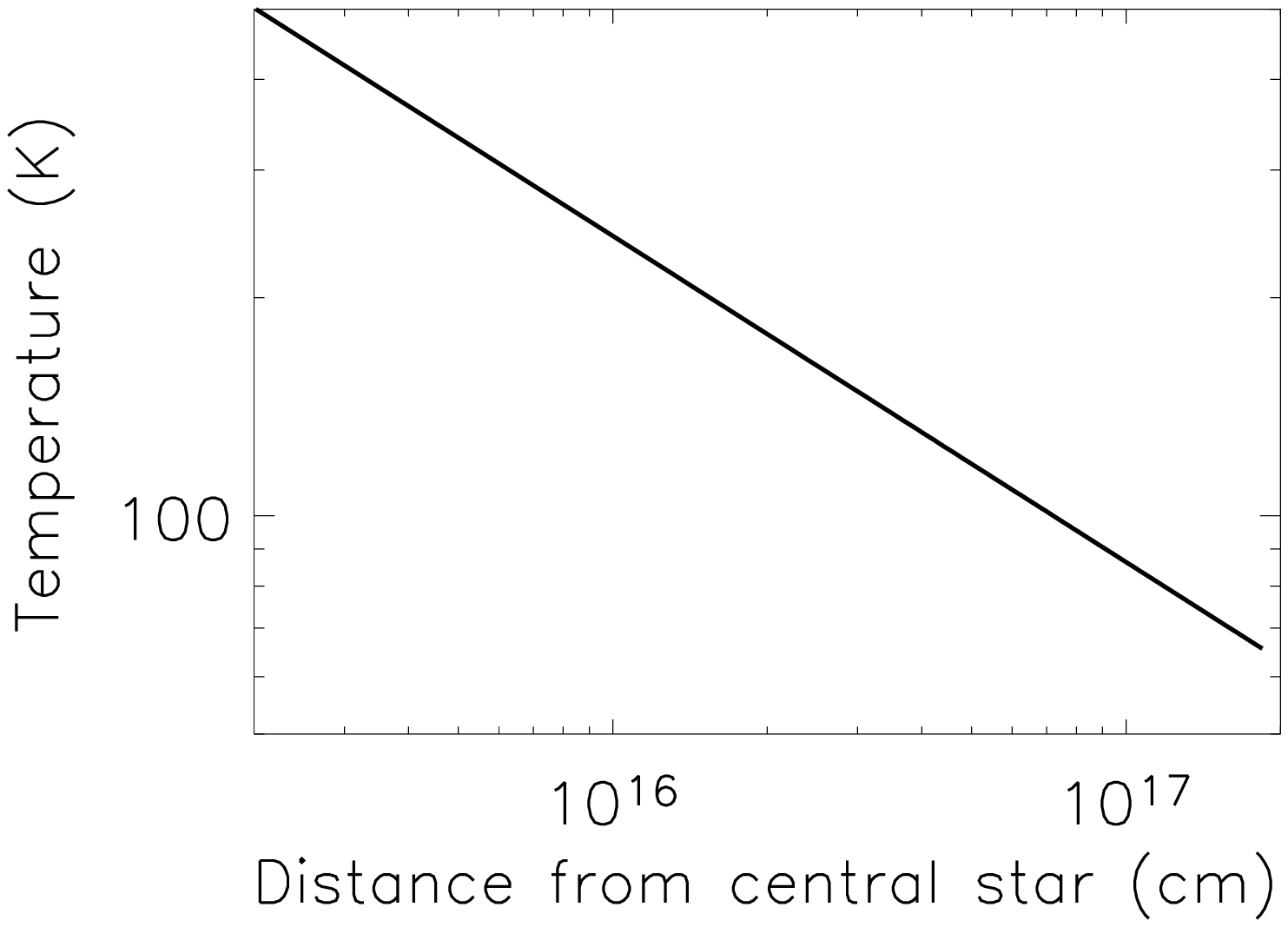}
\caption{The distribution of density ({\em left}) and temperature ({\em right}) as a function of position from the embedded source.}
\label{fig:embsou}
\end{minipage}
\end{figure}
\begin{figure}[h!]
\begin{minipage}[c]{0.5\textwidth}
\includegraphics[width=4.5cm]{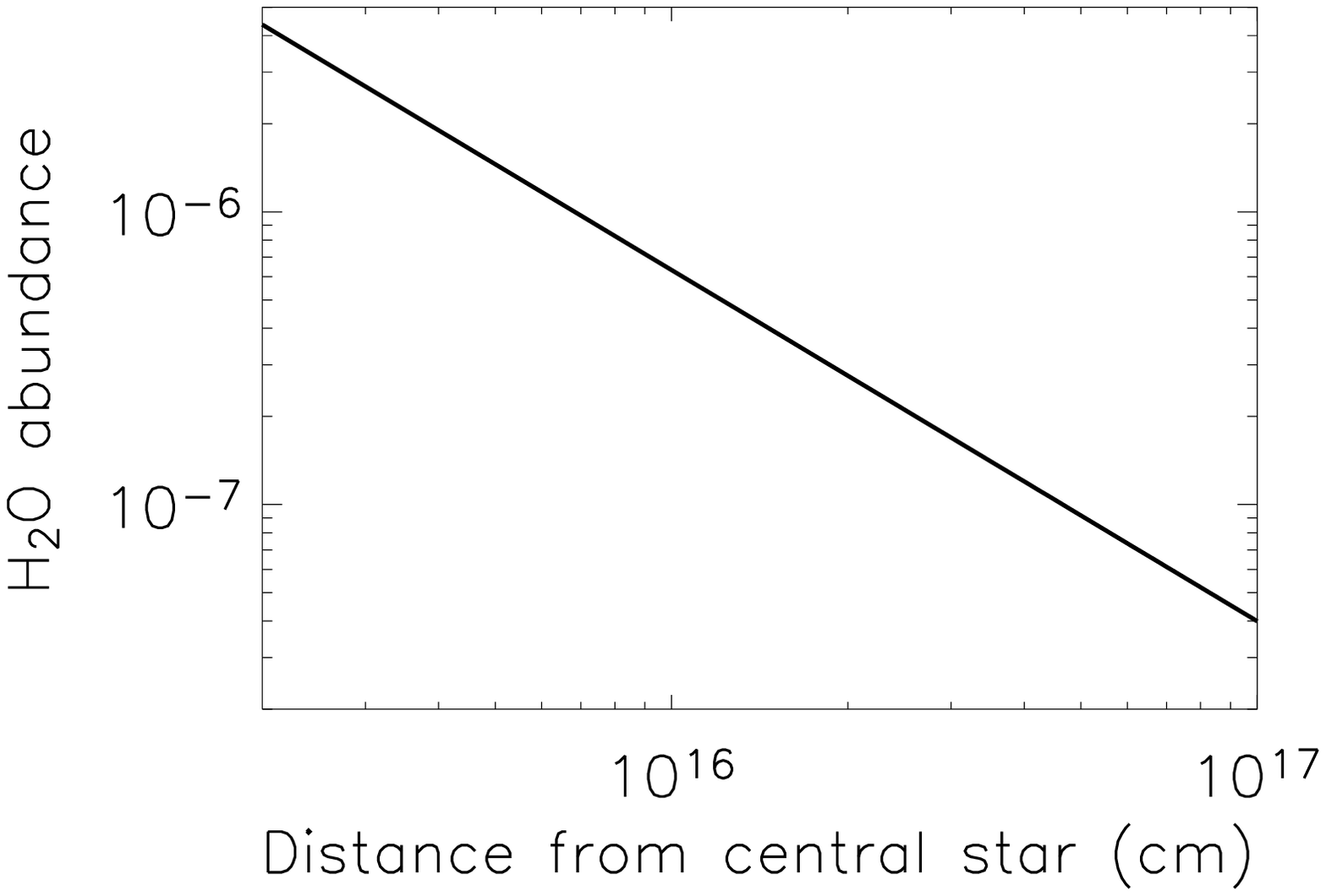}
\includegraphics[width=4.5cm]{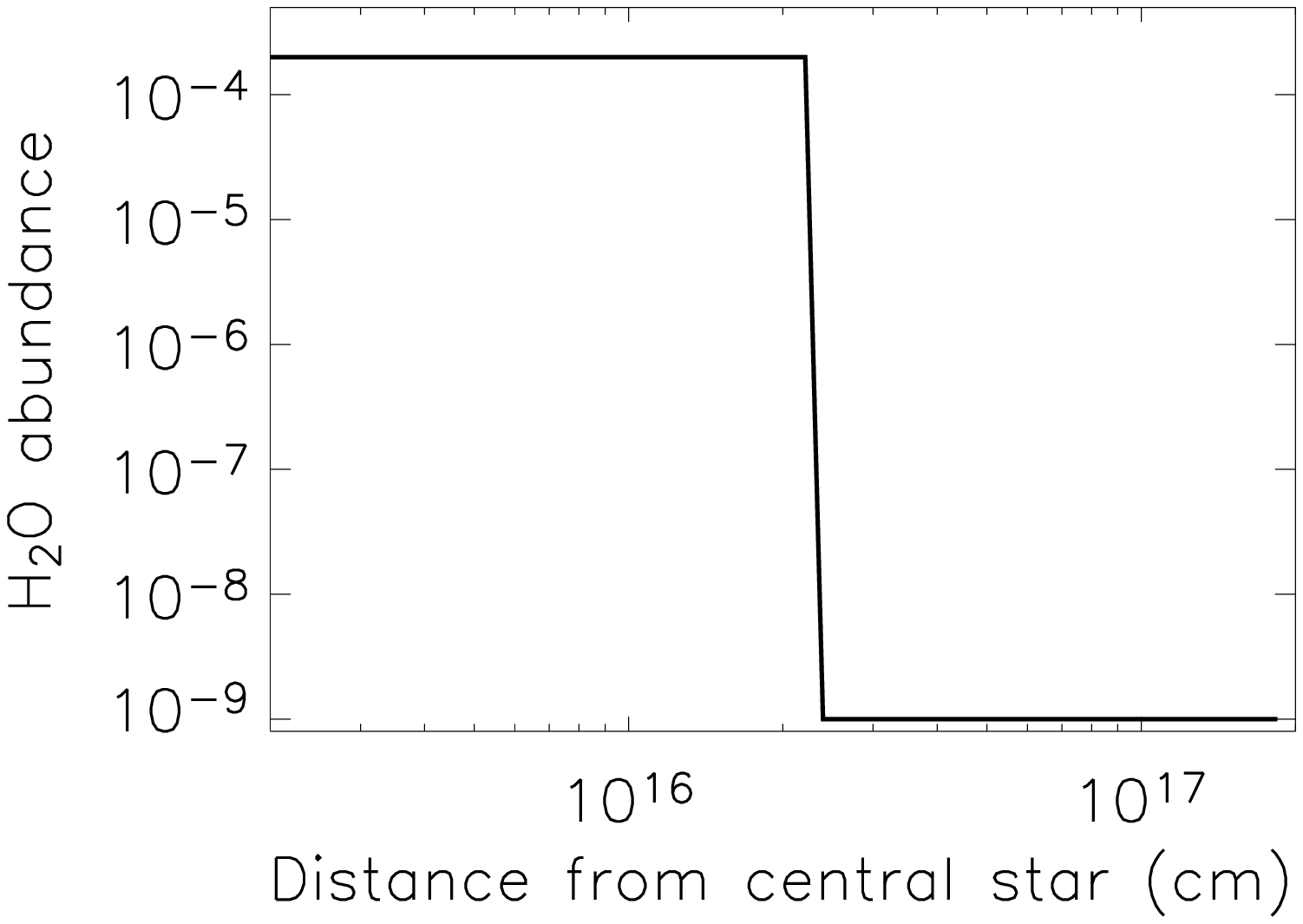}
\caption{The water abundance distribution, relative to \element[][]{H_2}, as a function of position from the embedded source for two (model I {\em (left)}; model II {\em (right)}) different scenarios.}
\label{fig:embsouabun}
\end{minipage}
\end{figure}
\begin{table}
\caption{Observed ISO line intensities $[$\mbox{$\mathrm{erg}\ \mathrm{cm}^{-2}\ \mathrm{s}^{-1}\ \mathrm{sr}^{-1}$}$]$ for S140 IRS1}
\begin{tabular}{cccc}
\hline 
\hline 
\multicolumn{4}{c}{Transition and Wavelength}\\
\hline
113.5 \mbox{$\mu$m} & 136.5 \mbox{$\mu$m} & 174.6 \mbox{$\mu$m} & 179.5 \mbox{$\mu$m} \\
$4_{14}$ $\rightarrow$ $3_{03}$ & $3_{30}$ $\rightarrow$ $3_{21}$ & $3_{03}$ $\rightarrow$ $2_{12}$ & $2_{12}$ $\rightarrow$ $1_{01}$ \\
\hline
$<$ 6.8 $\times$ ${10^{-6}}$ & $<$ 8.6 $\times$ ${10^{-6}}$ & 7.1 $\times$ ${10^{-5}}$ & $<$ 4.5 $\times$ ${10^{-5}}$\\
\hline
\end{tabular}
\label{tab:S140IRS1}
\end{table}
\begin{table*}
\caption{Embedded Source Model}
\begin{tabular}{ccccc}
\hline 
\hline 
  transition & Wavelength  & average & average & average \\
  & & intensity & intensity & intensity\\
  & & model I & model II & Poelman $\&$ Spaans (2005), PDR I\\
  & $[$\mbox{$\mu$m}$]$  & $[$\mbox{$\mathrm{erg}\ \mathrm{cm}^{-2}\ \mathrm{s}^{-1}\ \mathrm{sr}^{-1}$}$]$ & $[$\mbox{$\mathrm{erg}\ \mathrm{cm}^{-2}\ \mathrm{s}^{-1}\ \mathrm{sr}^{-1}$}$]$ & $[$\mbox{$\mathrm{erg}\ \mathrm{cm}^{-2}\ \mathrm{s}^{-1}\ \mathrm{sr}^{-1}$}$]$ \\
\hline
               o-\element[][]{H_2O} $1_{10}$ $\rightarrow$ $1_{01}$ & 538.3  & 5.0 $\times$ ${10^{-6}}$  & 2.2 $\times$ ${10^{-6}}$ & 1.1 $\times$ ${10^{-7}}$ \\ 
               o-\element[][]{H_2O} $2_{12}$ $\rightarrow$ $1_{01}$ & 179.5  & 4.5 $\times$ ${10^{-5}}$  & 1.4 $\times$ ${10^{-5}}$ & 6.2 $\times$ ${10^{-7}}$ \\ 
               o-\element[][]{H_2O} $2_{21}$ $\rightarrow$ $1_{10}$ & 108.0  & 3.4 $\times$ ${10^{-5}}$  & 1.4 $\times$ ${10^{-4}}$ & 5.7 $\times$ ${10^{-9}}$ \\ 
               o-\element[][]{H_2O} $2_{21}$ $\rightarrow$ $2_{12}$ & 180.5  & 1.1 $\times$ ${10^{-5}}$  & 3.8 $\times$ ${10^{-5}}$ & 4.1 $\times$ ${10^{-10}}$\\ 
               o-\element[][]{H_2O} $3_{03}$ $\rightarrow$ $2_{12}$ & 174.6  & 2.5 $\times$ ${10^{-5}}$  & 5.4 $\times$ ${10^{-5}}$ & 4.7 $\times$ ${10^{-9}}$ \\ 
               o-\element[][]{H_2O} $3_{12}$ $\rightarrow$ $2_{21}$ & 259.9  & 6.8 $\times$ ${10^{-6}}$  & 3.1 $\times$ ${10^{-5}}$ & 1.3 $\times$ ${10^{-11}}$\\ 
               o-\element[][]{H_2O} $3_{12}$ $\rightarrow$ $3_{03}$ & 273.2  & 4.9 $\times$ ${10^{-6}}$  & 1.6 $\times$ ${10^{-5}}$ & 7.8 $\times$ ${10^{-11}}$\\ 
               o-\element[][]{H_2O} $3_{21}$ $\rightarrow$ $2_{12}$ & 75.3   & 3.7 $\times$ ${10^{-5}}$  & 2.2 $\times$ ${10^{-4}}$ & -- \\ 
               o-\element[][]{H_2O} $3_{21}$ $\rightarrow$ $3_{12}$ & 257.7  & 2.7 $\times$ ${10^{-6}}$  & 8.0 $\times$ ${10^{-6}}$ & -- \\ 
               o-\element[][]{H_2O} $4_{14}$ $\rightarrow$ $3_{03}$ & 113.5  & 2.7 $\times$ ${10^{-5}}$  & 1.3 $\times$ ${10^{-4}}$ & -- \\ 
               o-\element[][]{H_2O} $3_{30}$ $\rightarrow$ $3_{21}$ & 136.5  & 5.0 $\times$ ${10^{-6}}$  & 2.2 $\times$ ${10^{-6}}$ & -- \\ 
\cline{1-5}
               p-\element[][]{H_2O} $1_{11}$ $\rightarrow$ $0_{00}$ & 269.3  & 1.5 $\times$ ${10^{-5}}$  & 1.5 $\times$ ${10^{-5}}$ & 2.5 $\times$ ${10^{-7}}$ \\ 
               p-\element[][]{H_2O} $2_{02}$ $\rightarrow$ $1_{11}$ & 303.4  & 1.2 $\times$ ${10^{-5}}$  & 1.0 $\times$ ${10^{-5}}$ & 2.7 $\times$ ${10^{-8}}$ \\ 
               p-\element[][]{H_2O} $2_{20}$ $\rightarrow$ $1_{11}$ & 100.9  & 1.7 $\times$ ${10^{-5}}$  & 1.4 $\times$ ${10^{-4}}$ & 8.8 $\times$ ${10^{-10}}$\\ 
               p-\element[][]{H_2O} $2_{11}$ $\rightarrow$ $2_{02}$ & 398.6  & 4.2 $\times$ ${10^{-6}}$  & 4.6 $\times$ ${10^{-6}}$ & 2.7 $\times$ ${10^{-10}}$\\ 
               p-\element[][]{H_2O} $2_{20}$ $\rightarrow$ $2_{11}$ & 243.9  & 2.2 $\times$ ${10^{-6}}$  & 1.4 $\times$ ${10^{-5}}$ & -- \\ 
               p-\element[][]{H_2O} $3_{13}$ $\rightarrow$ $2_{02}$ & 138.5  & 1.8 $\times$ ${10^{-5}}$  & 7.8 $\times$ ${10^{-5}}$ & -- \\ 
               p-\element[][]{H_2O} $3_{13}$ $\rightarrow$ $2_{20}^{\dagger}$ & 1635.7 & 1.9 $\times$ ${10^{-11}}$ & 1.6 $\times$ ${10^{-7}}$ & -- \\ 
               p-\element[][]{H_2O} $3_{22}$ $\rightarrow$ $2_{11}$ & 89.9   & 1.8 $\times$ ${10^{-5}}$  & 1.5 $\times$ ${10^{-4}}$ & -- \\
               p-\element[][]{H_2O} $4_{04}$ $\rightarrow$ $3_{13}$ & 125.3  & 1.4 $\times$ ${10^{-5}}$  & 9.5 $\times$ ${10^{-5}}$ & -- \\ 
\hline
\multicolumn{5}{l}{$^{\dagger}$ Like \cite{1997ApJ...489..122D} we find weak maser action in this line.}
\end{tabular}
\label{tab:ESM}
\end{table*}
Before going to Sect. \ref{sec:S140} in which we discuss the line profiles for the S140 region, we calculate the radiative transfer of certain transitions of ortho- and para-\element[][]{H_2O} when an embedded source is incorporated into the models of the S140 PDR presented in \cite{2005A&A...440..559P}. The PDR lies at the south-western edge of the molecular cloud \object{L1204}. The molecular cloud, which extends over more than 30$'$ (10 pc), lies at a distance $\sim$ 900 \mbox{pc}, and is illuminated by a BOV star HD211880, located at about $\sim$ 7$'$ to the cloud. As known from observations \citep[IRS 1 to 3,][]{1989ApJ...346..212E}, an embedded cluster of infrared sources is part of the S140 molecular cloud. The temperature and density structure characterizing these sources are different from the ones adopted in \cite{2005A&A...440..559P} for the PDR and the Extended Molecular Cloud (EMC), and is therefore a useful extension to the performed calculations in the aforementioned paper.\\
The calculations have been performed in 1D (201 gridpoints), with an adopted size of 0.06 \mbox{pc} and 0.12 \mbox{pc} for the embedded source in model I and II, respectively. Using spherical symmetry, the resulting level populations have been rotated to become a 3D solution from which the intensities have been computed. No systematic velocity field is included, but a velocity dispersion, $\Delta$$\mathrm{v_{es}}$ = ${(\mathrm{v}_\mathrm{th}^2 + \mathrm{v}_\mathrm{turb}^2)^{1/2}}$, of 2 \mbox{km $\mathrm{s}^{-1}$} is adopted. The temperature and density structures of the embedded source were derived by \cite{2000ApJ...537..283V, 2003A&A...412..133V}
, using a power-law density structure n = $n_0$$\mathrm{(r/r_0)^{-\alpha}}$. The adopted density and temperature profiles are plotted in Fig. \ref{fig:embsou}. Following \cite{2003A&A...406..937B}, our density increases from a few times $10^5$ \mbox{$\mathrm{cm}^{-3}$} at the edge to almost $10^8$ \mbox{$\mathrm{cm}^{-3}$} in the centre of the embedded source for both models. We assume, for simplicity, that the temperature of the gas is coupled to the dust because of the high densities. In model I and II, the temperature rises from, respectively, 80 \mbox{K} and 60 \mbox{K} at the edge towards 450 K in the centre. The corresponding \element[][]{H_2O} abundance profiles are plotted in Fig. \ref{fig:embsouabun}. Note that the average \element[][]{H_2O} abundance in the PDR is $\sim$ 1 $\times$ ${10^{-8}}$, whereas it is a factor 20 and 2 $\times$ $10^3$ higher in Model I and II, respectively. The adopted water abundance profiles are merely to illustrate the effects that are expected to result when water molecules are thermally released from warm dust grains \citep[c.f.][]{2003A&A...406..937B}. Note that our embedded sources are only $\sim$ 2 $\times$ $10^{17}$ \mbox{cm} in size, compared to $\sim$ $10^{18}$ cm in \cite{2003A&A...406..937B}. We feel that this is appropriate because $10^{17}$ \mbox{cm} is approximately the size of a clump in the PDR \citep{2005A&A...440..559P}. Hence, beyond $10^{17}$ \mbox{cm} photo-dissociation of water as described in \cite{2005A&A...440..559P} dominates.  Table \ref{tab:S140IRS1} lists the observed ISO line intensities discussed in \cite{2003A&A...406..937B}. The observed line fluxes $[$\mbox{$\mathrm{W}\ \mathrm{cm}^{-2}\ \mu\mathrm{m}^{-1}$}$]$ have been converted into line intensities $[$\mbox{$\mathrm{erg}\ \mathrm{cm}^{-2}\ \mathrm{s}^{-1}\ \mathrm{sr}^{-1}$}$]$ by adopting a beam size of $\sim$ 78$''$ ($\sim$ 0.4 \mbox{pc} at the distance of S140). The resulting intensities of the two adopted models are listed in Table \ref{tab:ESM}, together with the results found in \cite{2005A&A...440..559P} for the PDR without embedded sources. Higher densities and temperatures lead to increased intensities for high lying transitions. The $1_{10}$ $\rightarrow$ $1_{01}$ at 557 GHz is a factor $\sim$ 10 too high compared to the observed SWAS value for both the models. However, beam dilution effects are likely to alleviate this \citep{2001ApJ...548L.217S}. The predicted intensity for the $3_{03}$ $\rightarrow$ $2_{12}$ line at 174.6 \mbox{$\mu$m} is less than a factor of two lower than the observed value of 7.1 $\times$ $10^{-5}$ \mbox{$\mathrm{erg}\ \mathrm{cm}^{-2}\ \mathrm{s}^{-1}\ \mathrm{sr}^{-1}$} for model II, while model I is lower by a factor of $\sim$ 3. A core of water, as in model II, is preferred to model this transition. The predicted intensities for the $2_{12}$ $\rightarrow$ $1_{01}$ line at 179.5 \mbox{$\mu$m} and the $3_{30}$ $\rightarrow$ $3_{21}$ line at 136.5 \mbox{$\mu$m} are in good agreement with the observed upper limits for both models. The observed upper limit for the $4_{14}$ $\rightarrow$ $3_{03}$ line at 113.5 \mbox{$\mu$m} is best reproduced by model I. Model II produces too much intensity for this transition by a factor of $\sim$ 20.\\
The intensities listed in Table \ref{tab:ESM} show that water emission from a PDR source, like S140, requires a combination of a pure PDR contribution for o-\element[][]{H_2O} $1_{10}$ $\rightarrow$ $1_{01}$ \citep{2005A&A...440..559P}, and an embedded source for high excitation lines (Model I/II) in order to match the observations.
\section{\label{sec:S140} Line profiles}
\begin{figure}[h!]
\begin{minipage}[c]{0.5\textwidth}
\includegraphics[width=4.0cm]{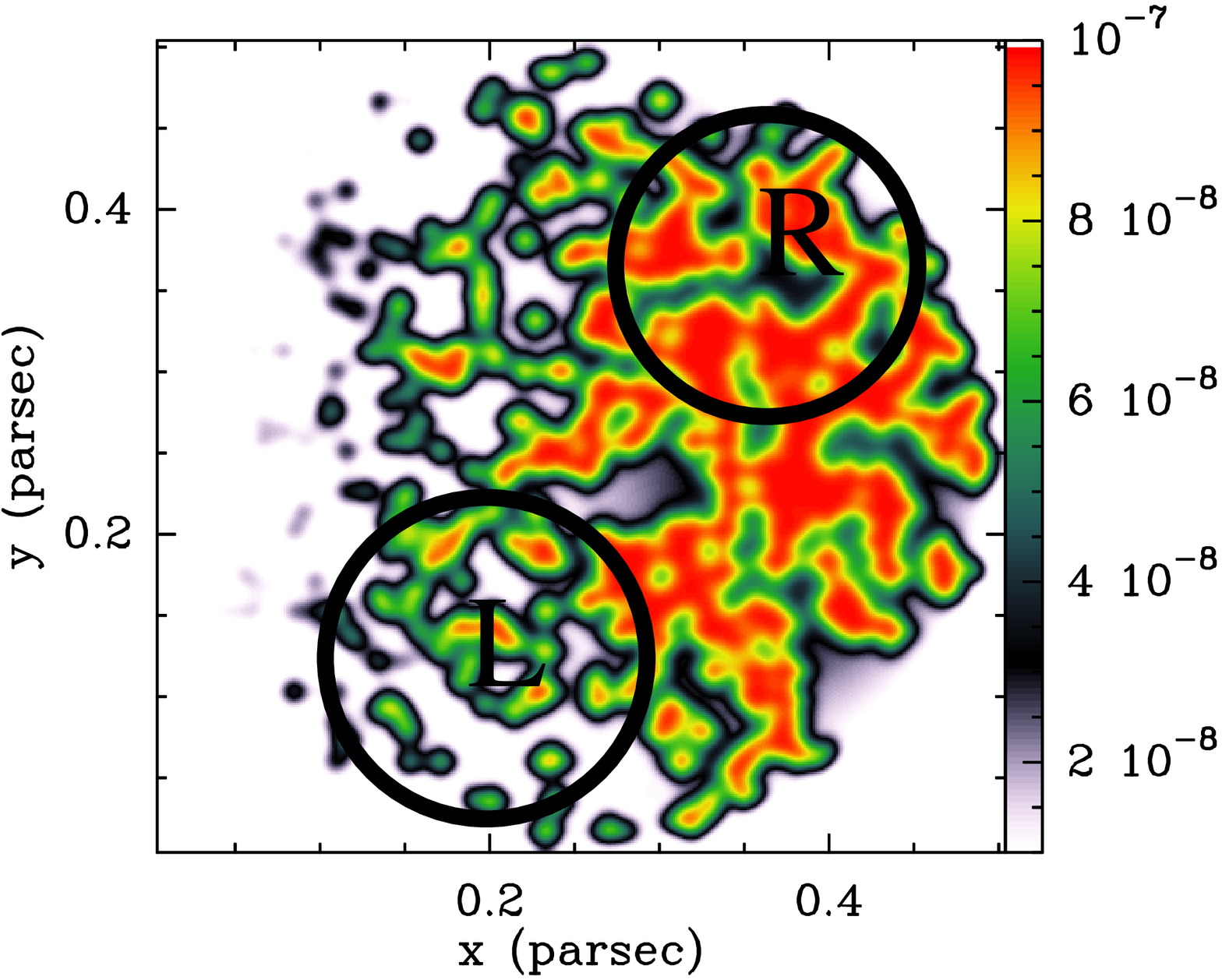} 
\includegraphics[width=4.2cm]{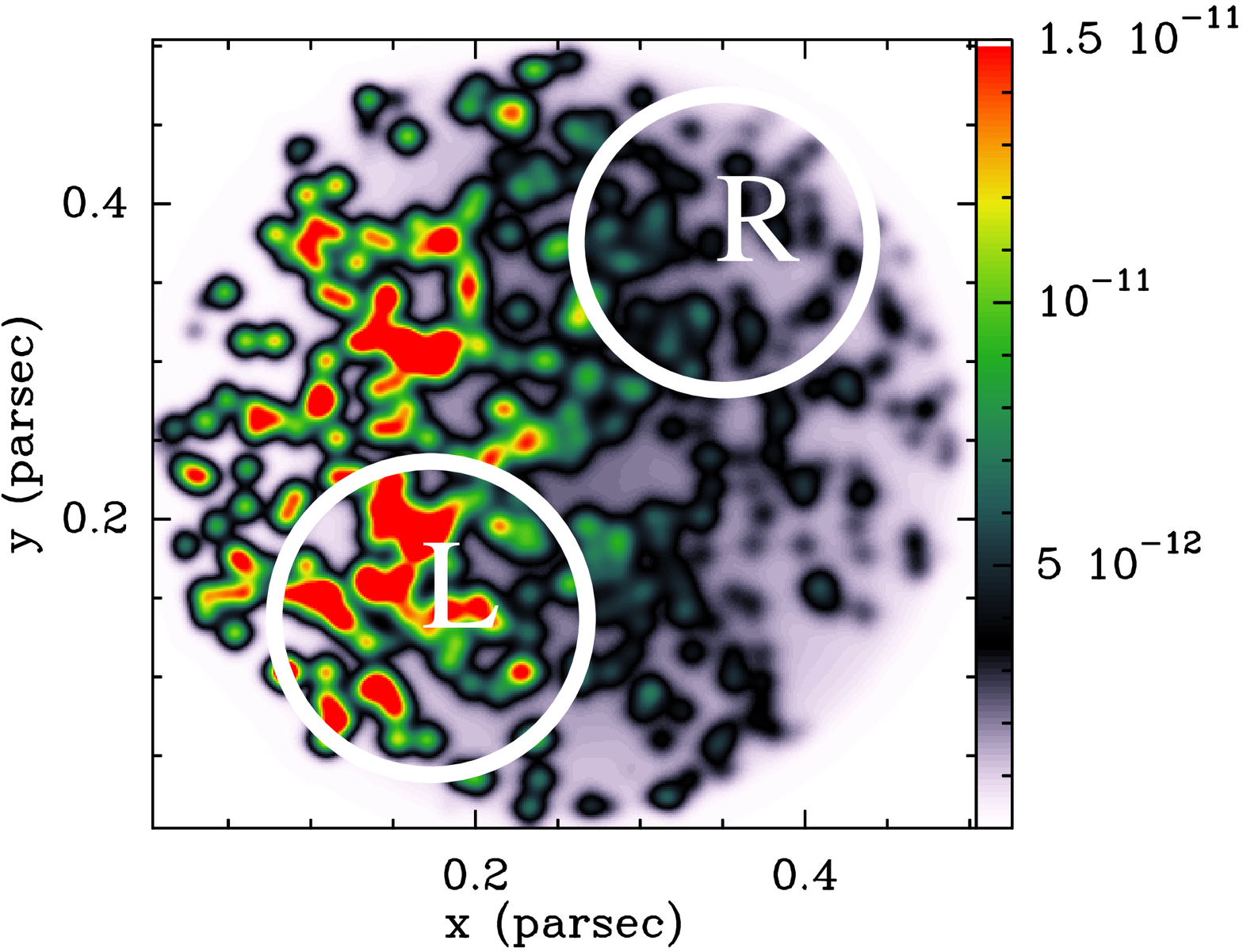}  
\caption{The figures show the predicted distribution of the intensities (erg $\mathrm{cm^{-2}}$ $\mathrm{s^{-1}}$ $\mathrm{sr^{-1}}$) of two transitions of ortho-$\mathrm{H_2O}$. The left figure depicts $\mathrm{1_{10}}$ $\rightarrow$ $\mathrm{1_{01}}$, on the right one has the $\mathrm{3_{12}}\ \rightarrow\ \mathrm{2_{21}}$ transition. The star is located to the left of the figures.}
\label{fig:S140}
\end{minipage}
\end{figure}
\begin{table}[h!]
\begin{minipage}[b]{\columnwidth}
\renewcommand{\footnoterule}{}
\caption{PDR model parameters}
\label{tab:parameters2}
\centering
\begin{tabular}{lccccccc}
\hline\hline  
Model & size & ${n_\mathrm{c}}$\footnote{ Clump density} & $n_\mathrm{ic}$\footnote{ Interclump density} & $F$\footnote{ Fraction of the total volume that is occupied by the clumps}& $l_\mathrm{c}$\footnote{ Clump size} & $\Delta$$\mathrm{v}_\mathrm{d}$\footnote{ Velocity dispersion} \\
 & $[$\mbox{pc}$]$ & $[$\mbox{$\mathrm{cm}^{-3}$}$]$ & $[$\mbox{$\mathrm{cm}^{-3}$}$]$& $[$\% $]$ & $[$\mbox{pc}$]$ &$[$\mbox{$\mathrm{km}\ \mathrm{s}^{-1}$}$]$\\
\hline
PDR & 0.5 & 4 $\times$ ${10^5}$ & 1 $\times$ ${10^4}$ & 2.5 & 0.03  & 1.2\\\hline
\end{tabular}
\end{minipage}
\end{table}
\begin{figure*}[h!]
\includegraphics[width=5.7cm]{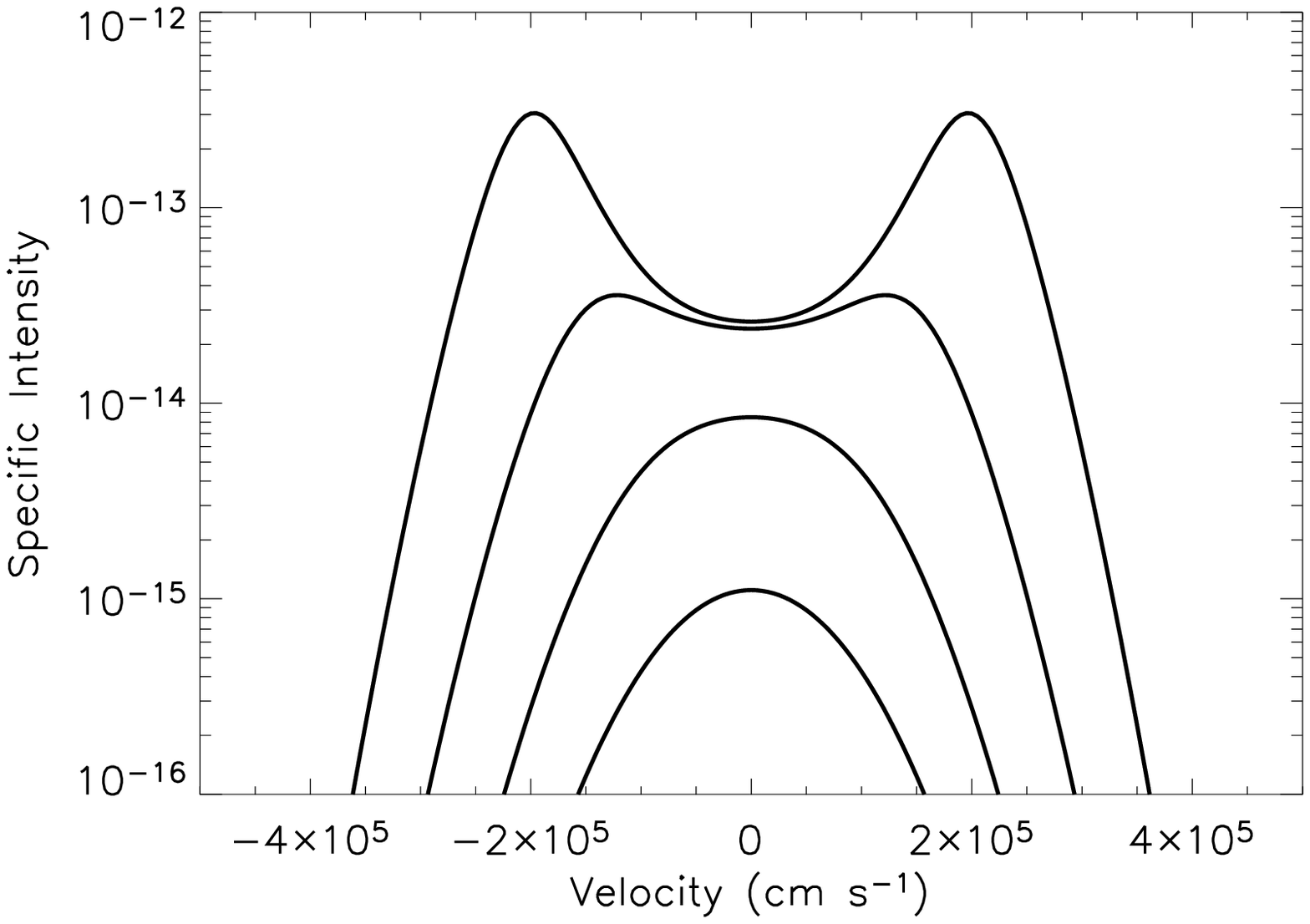}
\includegraphics[width=5.7cm]{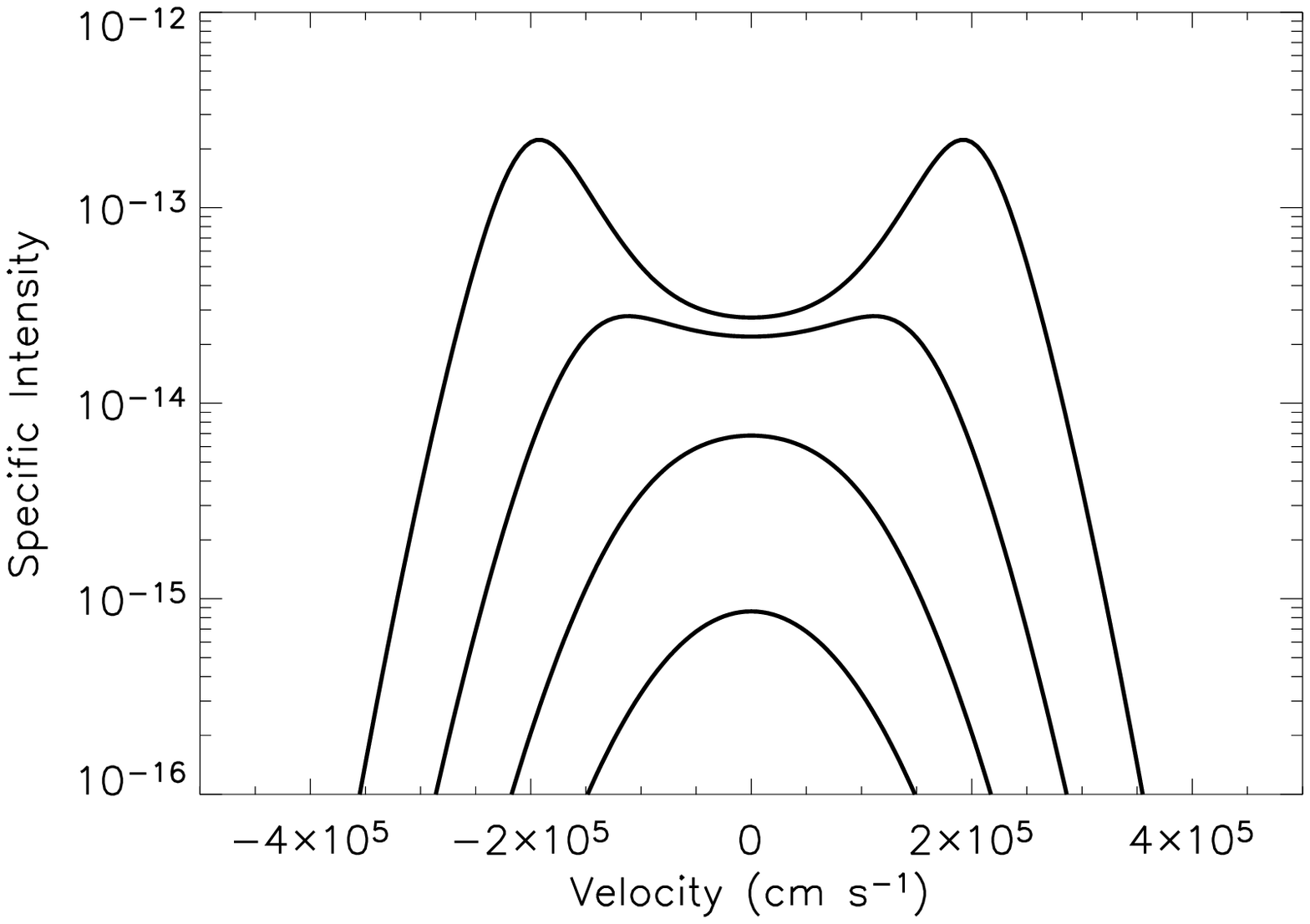}
\includegraphics[width=5.7cm]{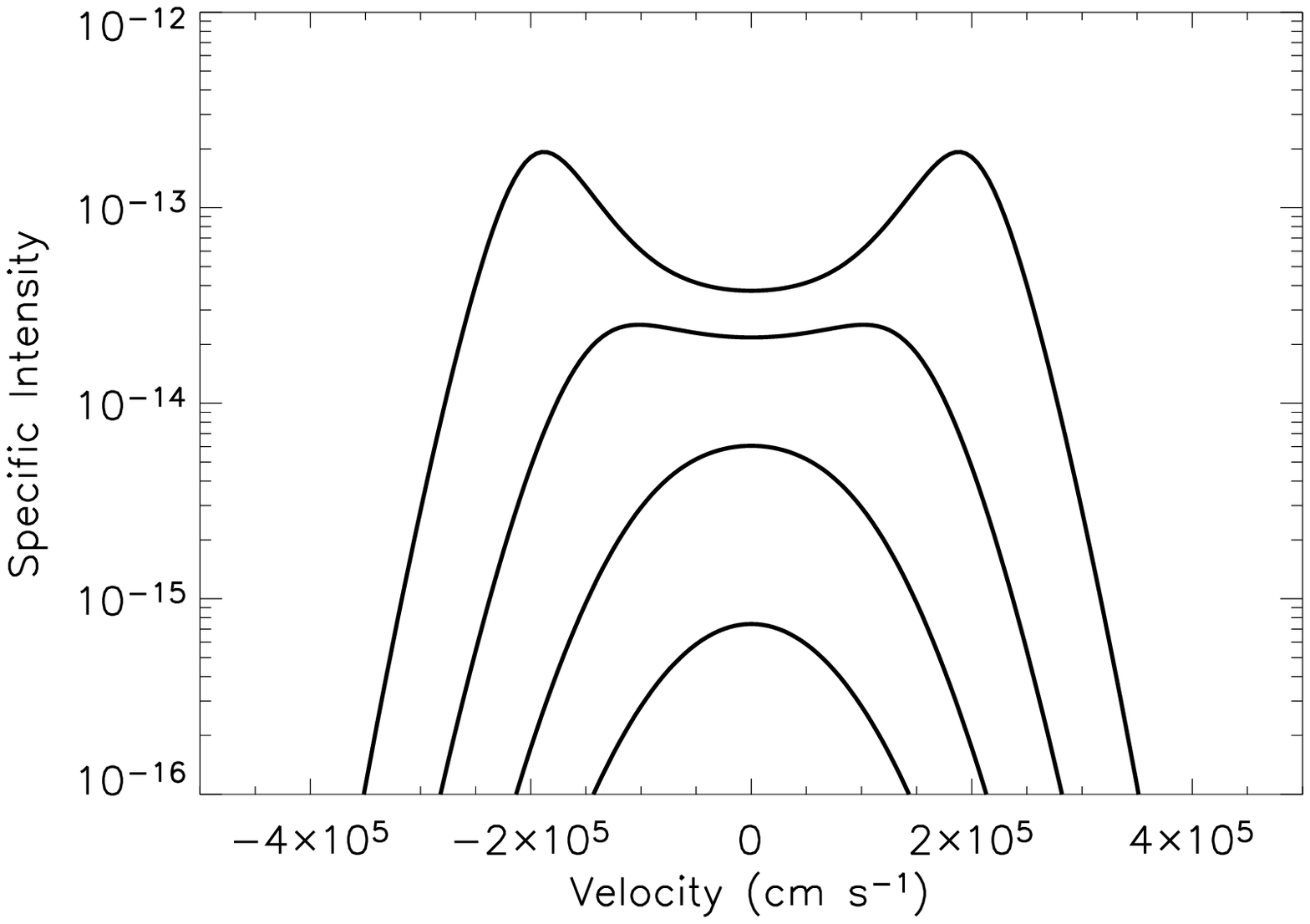}
\caption{Line profiles for a static two-level problem ($\mathrm{1_{01}\ and\ 1_{01}\ states}$). The line profiles are calculated for a 1D ({\em left}), 2D ({\em middle}) and 3D ({\em right}) configuration for different o-\element[][]{H_2O} abundances. The bottom curve resembles the line profile when the o-\element[][]{H_2O} abundance is $10^{-10}$, whereas the top one is the line profile in case the o-\element[][]{H_2O} abundance is $10^{-7}$. Y-axis in units of \mbox{$\mathrm{erg}\ \mathrm{cm}^{-2}\ \mathrm{s}^{-1}\ \mathrm{sr}^{-1}$} per \mbox{$\mathrm{cm}\ \mathrm{s^{-1}}$}.}
\label{fig:benchLP}
\end{figure*}
\begin{figure*}[h!]
\includegraphics[width=5.7cm]{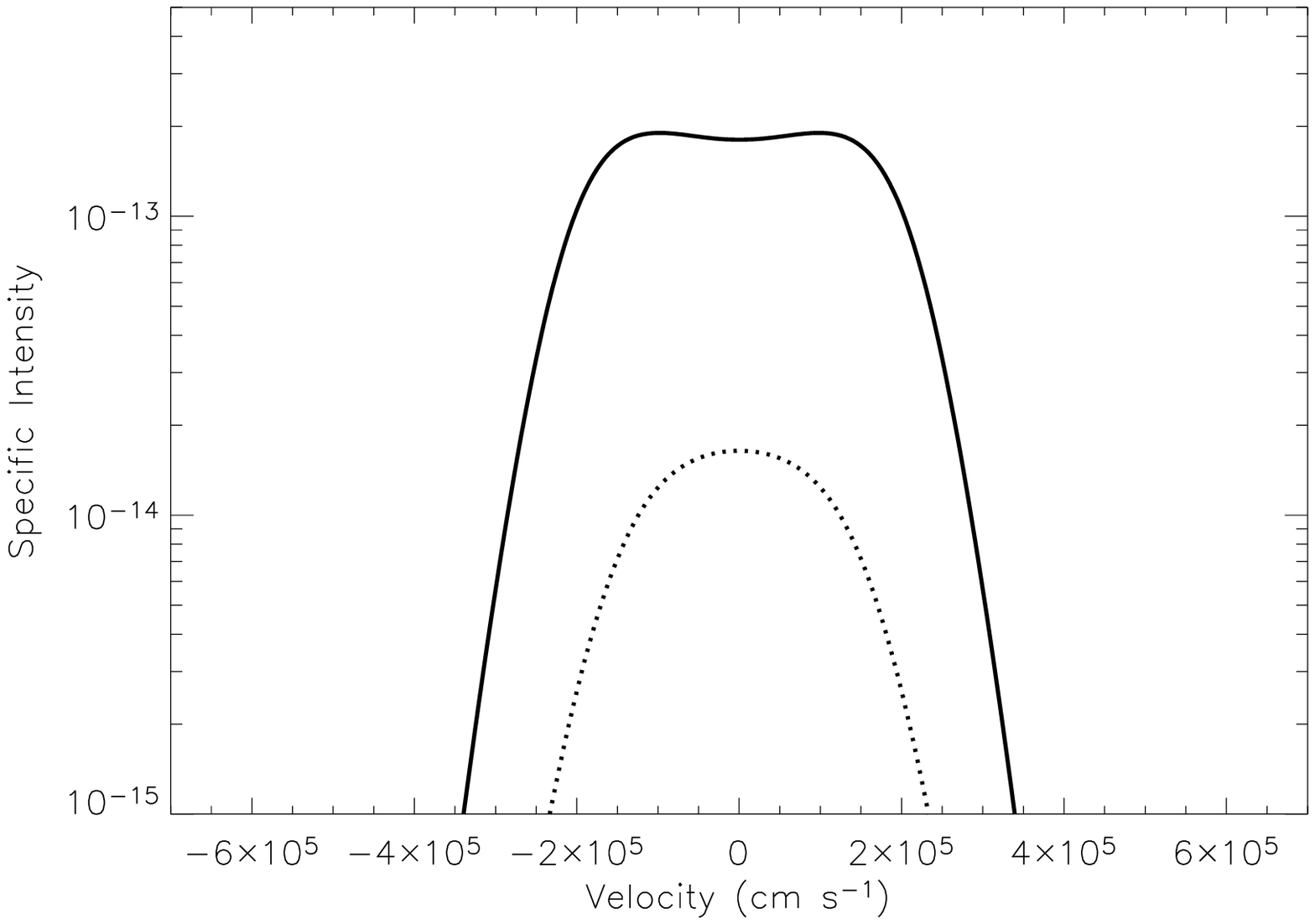}
\includegraphics[width=5.7cm]{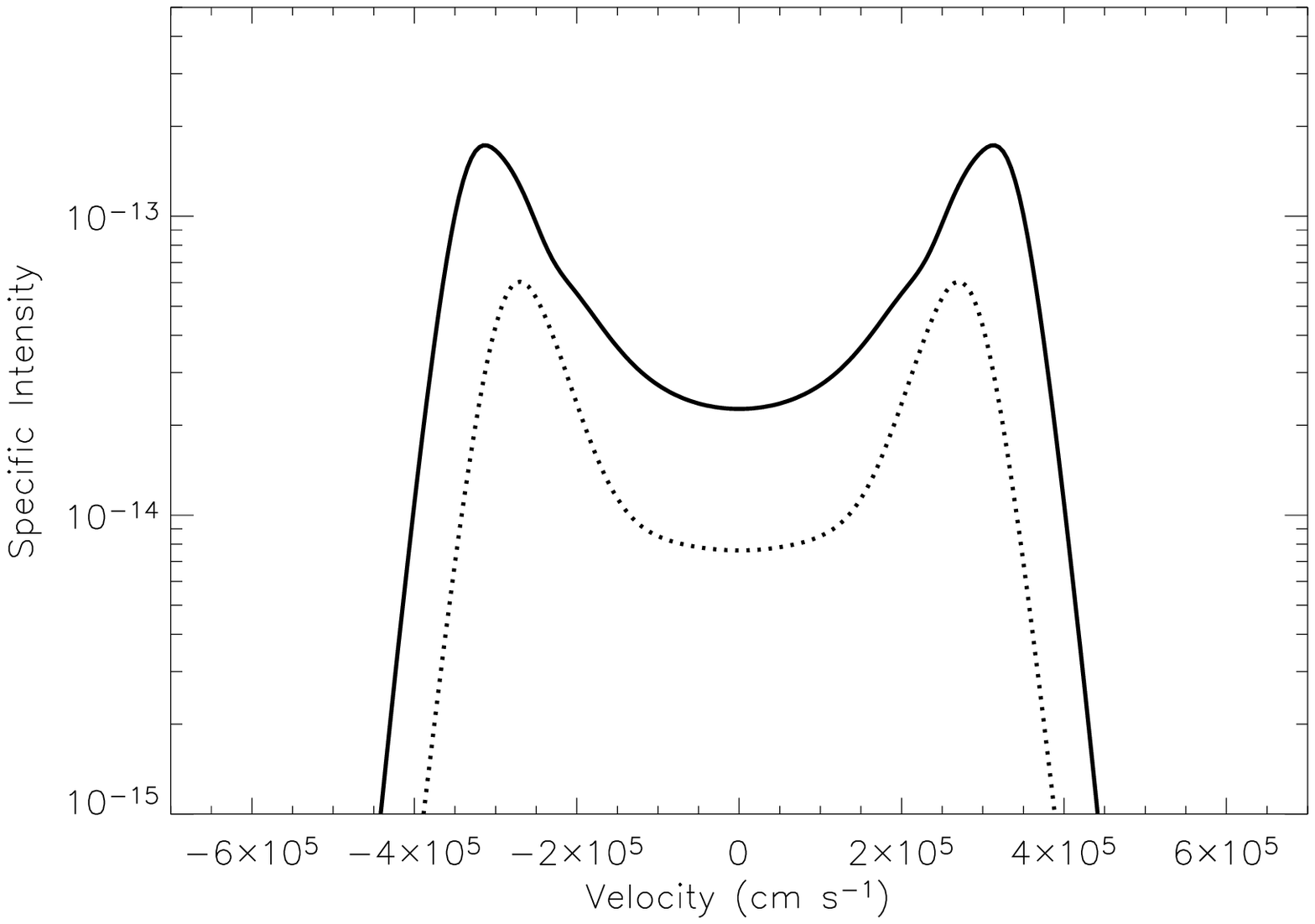}
\includegraphics[width=5.7cm]{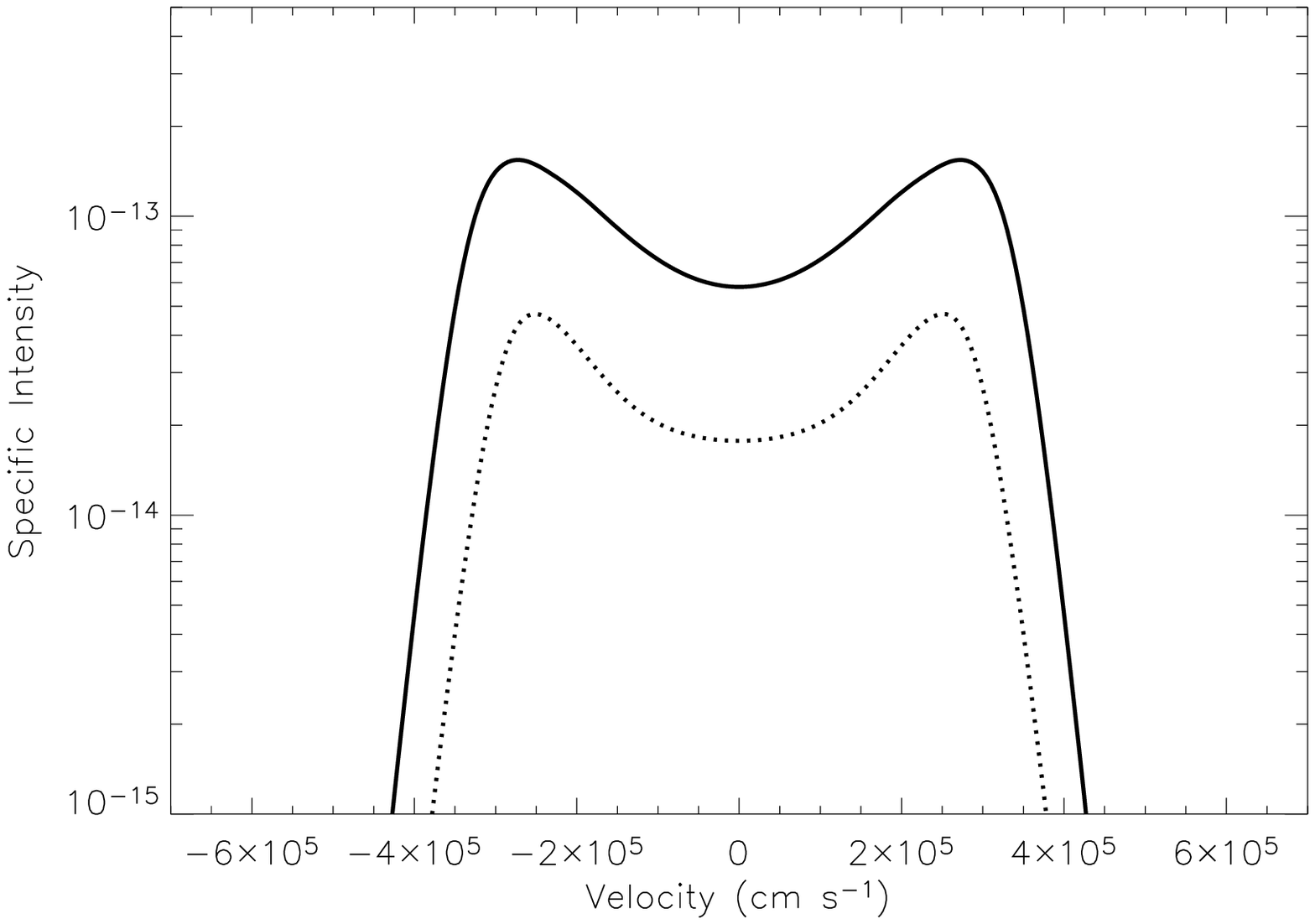}
\caption{Line profiles for the $\mathrm{1_{10}}$ $\rightarrow$ $\mathrm{1_{01}}$ transition in case of a static PDR are plotted. Line profiles are plotted when zooming in on a certain area in the PDR (left and middle panel), and when taking the mean over the PDR (right panel). The left ({\em middle}) panel represents the south-west ({\em north-east}) part of the cloud. Dotted ({\em solid}) lines are the line profiles in case of an ({\em in})homogeneous model. Y-axis in units of \mbox{$\mathrm{erg}\ \mathrm{cm}^{-2}\ \mathrm{s}^{-1}\ \mathrm{sr}^{-1}$} per \mbox{$\mathrm{cm}\ \mathrm{s^{-1}}$}.}
\label{fig:S140LPstatic}
\end{figure*}
\begin{figure}[h!]
\includegraphics[width=5.7cm]{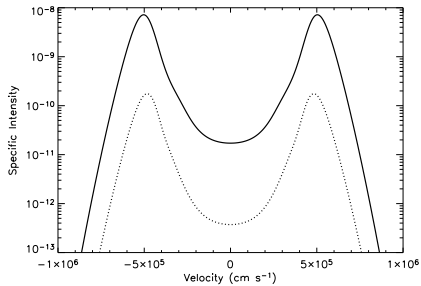}
\caption{Line profiles for the $1_{10}$ $\rightarrow$ $1_{01}$ ({\em dashed}) and  $3_{03}$ $\rightarrow$ $2_{12}$ ({\em solid}) transitions of o-\element[][]{H_2O} for a static embedded source ({\em Model I}). Y-axis in units of \mbox{$\mathrm{erg}\ \mathrm{cm}^{-2}\ \mathrm{s}^{-1}\ \mathrm{sr}^{-1}$} per \mbox{$\mathrm{cm}\ \mathrm{s^{-1}}$}.}
\label{fig:embeddedsourcelineprofile}
\end{figure}
\begin{figure*}[h!]
\includegraphics[width=5.7cm]{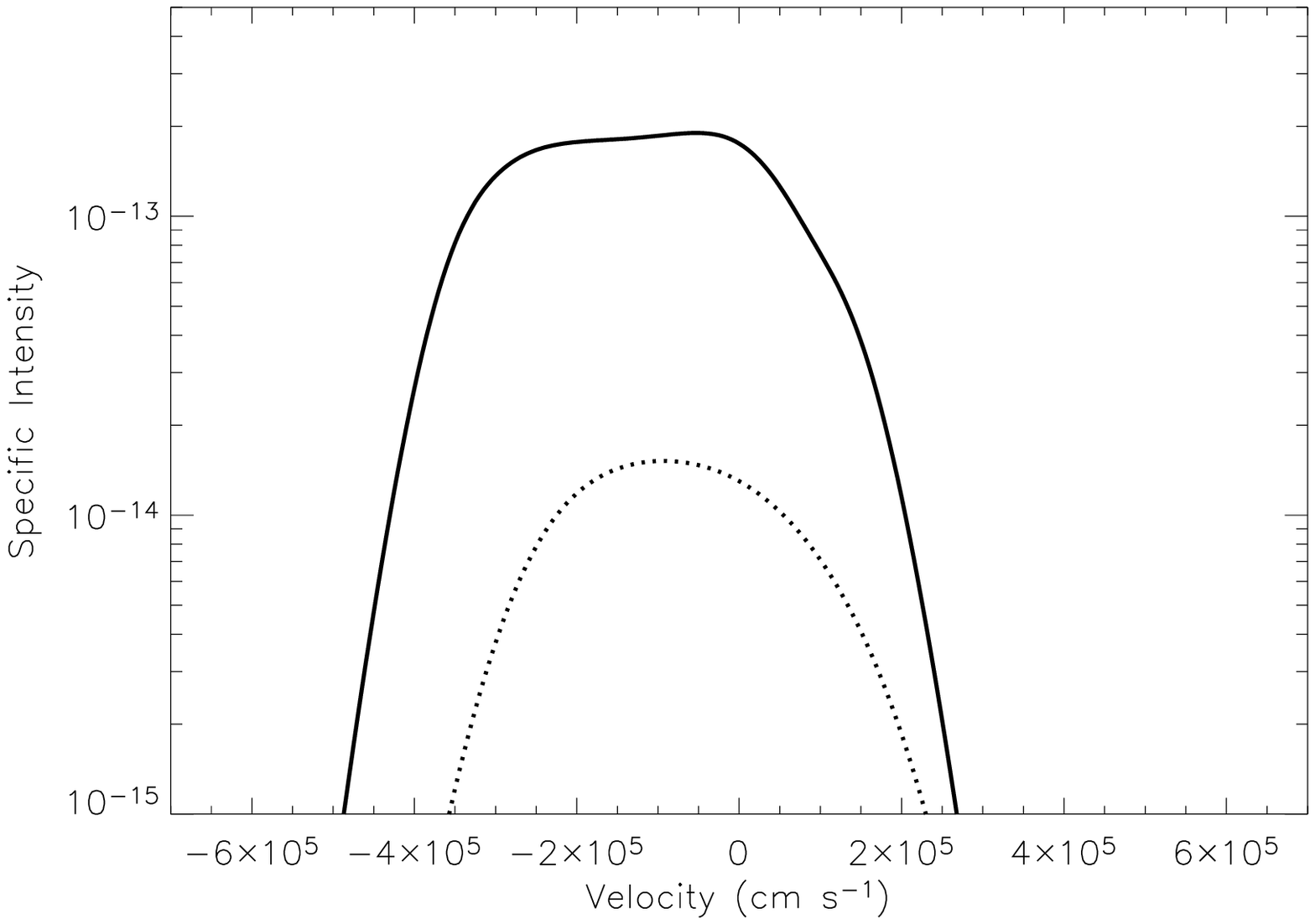}
\includegraphics[width=5.7cm]{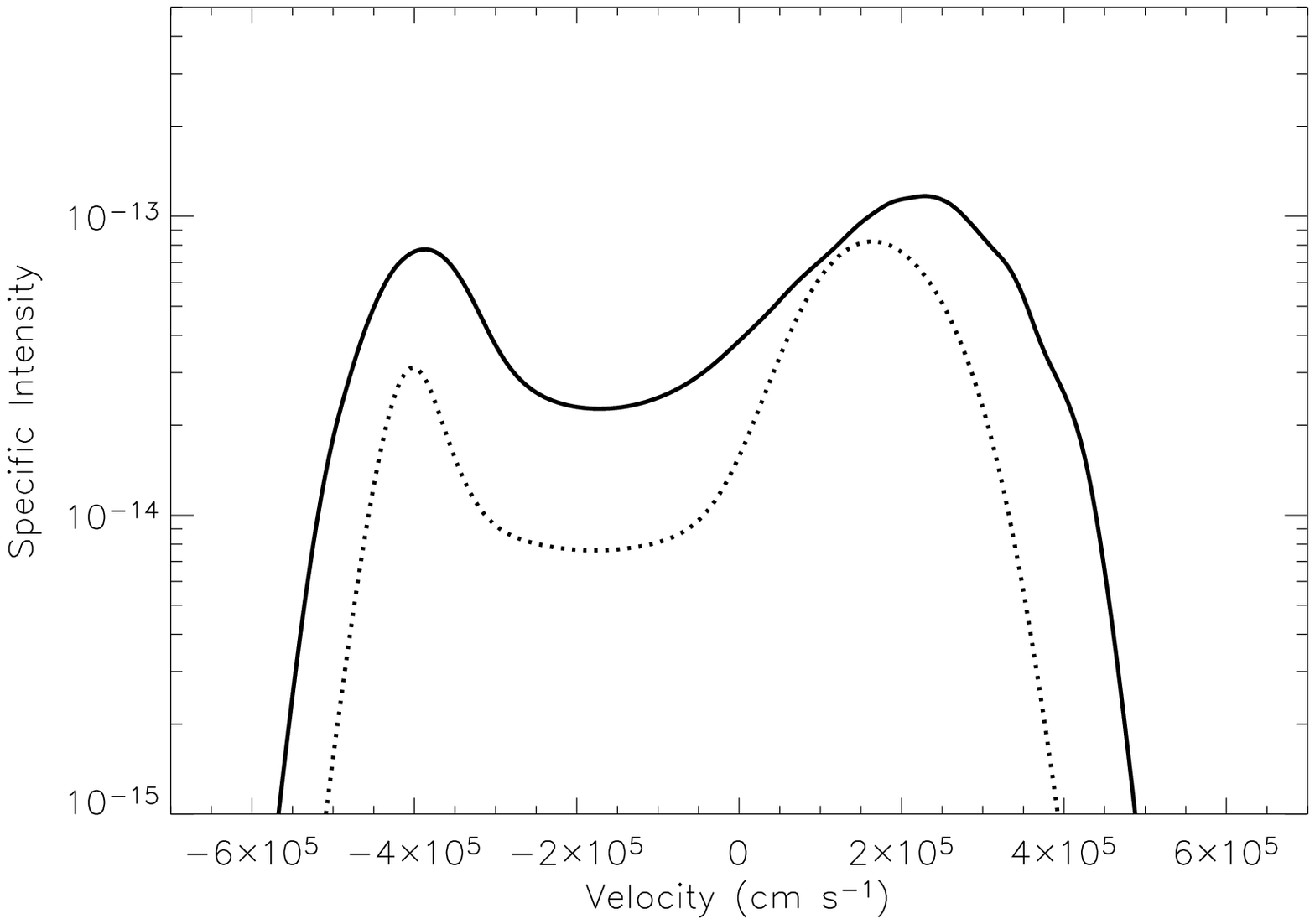}
\includegraphics[width=5.7cm]{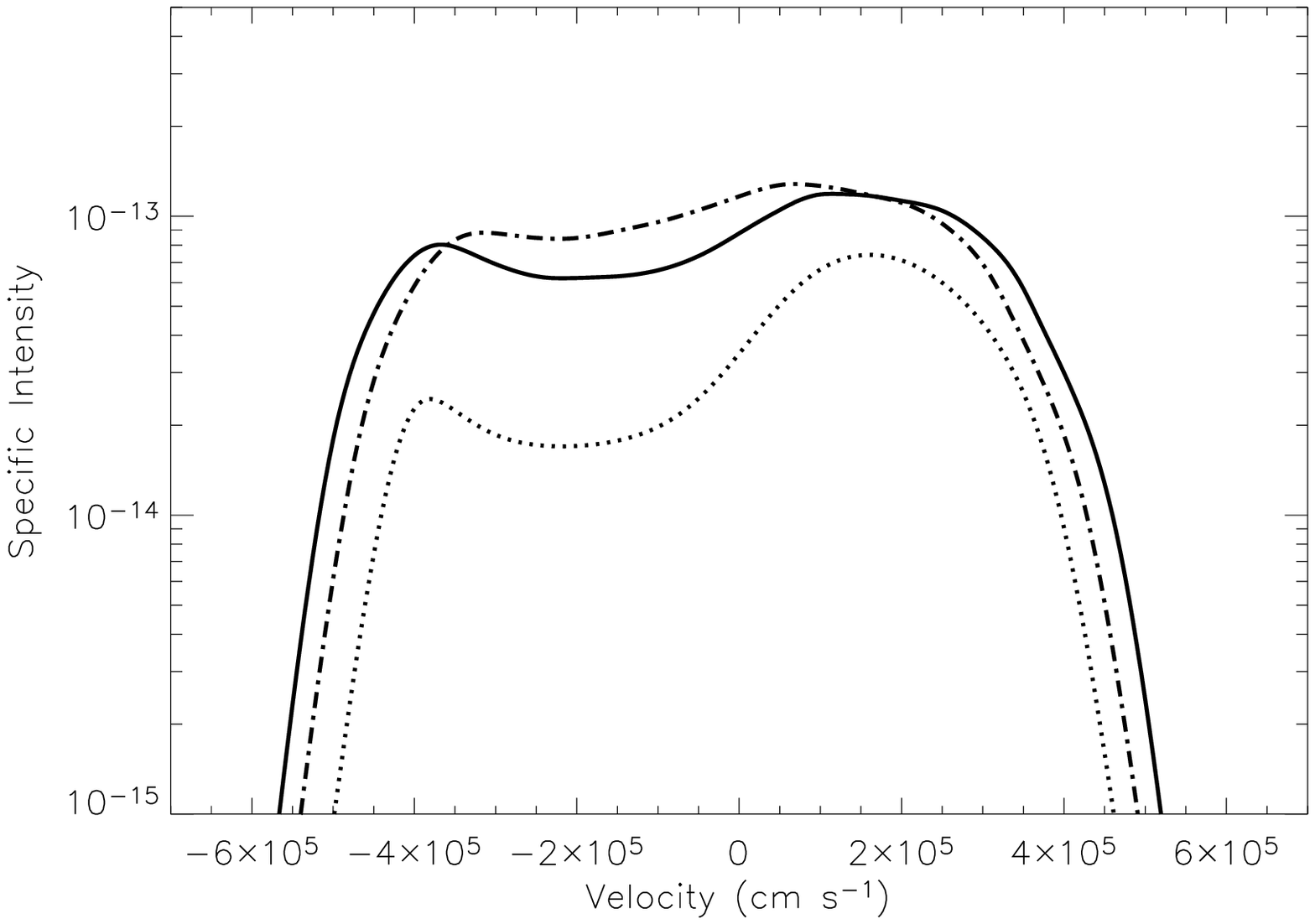}
\caption{Line profiles for the $\mathrm{1_{10}}$ $\rightarrow$ $\mathrm{1_{01}}$ transition are plotted when a systematic outflow velocity field is employed. Line profiles are plotted when zooming in on a certain area in the PDR (left and middle panel), and when taking the mean over the PDR (right panel). The left ({\em middle}) panel represents the south-west ({\em north-east}) part of the PDR. Dotted ({\em solid}) lines are the line profiles in case of an ({\em in})homogeneous model. In the right panel a line profile for a model with a smaller clump size, $l_c$ = 0.02 \mbox{pc}, (dashed-dotted) is also plotted. Y-axis in units of \mbox{$\mathrm{erg}\ \mathrm{cm}^{-2}\ \mathrm{s}^{-1}\ \mathrm{sr}^{-1}$} per \mbox{$\mathrm{cm}\ \mathrm{s^{-1}}$}.}
\label{fig:S140LPoutflow}
\end{figure*}
In this section, we present line profiles for the $1_{10}$ $\rightarrow$ $1_{01}$ transition of ortho-\element[][]{H_2O} at a few positions in the S140 PDR (without embedded sources), as well as line profiles when the beam is as large as the PDR (Fig. \ref{fig:S140LPstatic} $\&$ \ref{fig:S140LPoutflow}). Line profiles for a static embedded source (Fig. \ref{fig:embeddedsourcelineprofile}) are also shown.\\
The computed temperature, density and water abundance distributions, as calculated with the inhomogeneous code of Spaans (1996), is used as input for solving the \element[][]{H_2O} equations of statistical equilibrium by means of an escape probability method. For a full description of the model, we refer to \cite{2005A&A...440..559P}. Fig. \ref{fig:S140} shows illustrative maps of the predicted distribution of the intensities of two ortho-\element[][]{H_2O} transitions. The parameters used for this model are listed in Table \ref{tab:parameters2} (see \cite{2005A&A...440..559P}). It is seen that different lines probe different regions in the PDR. The $\mathrm{1_{10}}$ $\rightarrow$ $\mathrm{1_{01}}$ transition peaks near the eastern edge of the PDR where the gas temperature drops towards 50 K, the $\mathrm{3_{12}}$ $\rightarrow$ $\mathrm{2_{21}}$ line peaks where the temperatures are highest, at the western edge of the cloud. In the following we present line profiles when zooming in on different regions of the PDR, see Fig. \ref{fig:S140}. The two areas, L and R, are indicated by a circle.\\
In Figures \ref{fig:S140LPstatic} $\&$ \ref{fig:S140LPoutflow} we plot the line profiles for a static PDR model, and a PDR model with an outflow velocity field, respectively. Fig. \ref{fig:S140LPstatic} shows the line profiles when S140 is considered as a spherical, static cloud. A homogenous ({\em dotted}) and inhomogenous ({\em solid}) distribution have been considered. The left ({\em middle}) panel represents the line profiles when one takes the mean over a region, with a size comparable to the HIFI beam$\footnote{Half Power Beamwidth between 13$''$ and 39$''$, depending on frequency band.}$, toward the south-west ({\em north-east}) side of the cloud. It is seen that different regions lead to different line shapes. In the left part of the cloud the abundance of water is lower, due to photo-dissociation, compared to the other side of the cloud. This leads to a lower water column, hence lower opacity when compared to the right side of the cloud. Therefore, a single-peaked line profile arises in the left part, whereas the line profile in the right part is self-absorbed due to opacity effects. In the inhomogeneous case, the clumps are the opacity source, increasing the opacity in the left part, resulting in a top-flatted profile. 
The right panel of Figure \ref{fig:S140LPstatic} gives the line profiles when we take the mean over the PDR in its totallity. The shape of the curves is similar in both cases. It is seen that the inhomogeneous distribution produces a higher intensity than the homogeneous distribution.\\
In Fig. \ref{fig:embeddedsourcelineprofile} we plot the line profiles for the o-\element[][]{H_2O} $1_{10}$ $\rightarrow$ $1_{01}$ and  $3_{03}$ $\rightarrow$ $2_{12}$ transitions arising from the static embedded source described in Section \ref{sec:embsou}. One notices the difference with Fig. \ref{fig:S140LPstatic}, in that the larger opacity in the lines causes the profiles to be very strongly self-absorbed. Inclusion of a velocity gradient with magnitude $\sim$ 3 $\Delta$$\mathrm{v_{es}}$ does not remove this self-absorption characteristic.\\ 
Fig. \ref{fig:S140LPoutflow} shows the line profiles for a PDR model with an outflow velocity field. We consider a velocity field, $\mathrm{V_{outflow}}$, ranging from -2 to +2 \mbox{$\mathrm{km}\ \mathrm{s}^{-1}$}, yielding a velocity equal to zero in the middle of the cloud, and gradually increasing towards the edge of the cloud, i.e., $\mathrm{V_{outflow}}$ $\sim$ R. As in Fig. \ref{fig:S140LPstatic}, a homogenous ({\em dotted}) and inhomogenous ({\em solid}) density distribution have been considered. The left ({\em middle}) panel represents the line profiles when one takes the mean over a region, with a size comparable to the HIFI beam, toward the south-west edge ({\em north-east}) side of the cloud. Line profiles coming from the right part have a larger width than the ones from the left part. The overall tendency of the shapes of the line profiles in the inhomogeneous case is comparable to the homogeneous models. At scales as large as the PDR, (right panel of Fig. \ref{fig:S140LPoutflow}) line profiles are plotted in case of a homogeneous ({\em dotted}) and inhomogeneous ({\em solid}) PDR model. Also, a line profile for a model with a smaller clump size $l_c$ = 0.02 \mbox{pc} (dashed-dotted), i.e., with clumps that are about 30 $\%$ smaller, is plotted. This clump size is consistent with other PDR characteristics \citep{1997A&A...323..953S} of S140. We see that the line profile for the homogeneous model is more self-absorbed than when an inhomogeneous distribution is adopted. Reducing the clump size by a factor of 1.5, leads to a more single-peaked line profile, comparable with observations \citep{2000ApJ...539L.101S, 2000ApJ...539L.119A}. Although not shown, altering the velocity dispersion to 2 \mbox{$\mathrm{km}\ \mathrm{s}^{-1}$} broadens the line profile by a factor of $\sim$ 2, in disagreement with observations.\\
If only microturbulence is considered within the cloud, then stronger self-absorption effects occur than when macroturbulent motions are considered as well. The widths of the profiles are broader in the eastern part of the cloud, compared to the western side. However, the peak intensity is higher in the west side of the cloud. Note that the slopes of the line profiles depend on the adopted velocity dispersion. The differences between the slopes of the homogeneous and inhomogeneous models do not only depend on the density distribution (clumpiness versus homogeneity), but are also a result of the chemical and temperature balance within the clouds. Velocity fields with a moderate velocity gradient, as in $\mathrm{V_{outflow}}$, give rise to an almost flat line profile with a FWHM of $\sim$ 6 \mbox{$\mathrm{km}\ \mathrm{s}^{-1}$}, in agreement with observations \citep{2000ApJ...539L.101S, 2000ApJ...539L.119A}. Similar results are found for an infall velocity field with a comparable gradient.
\section{\label{sec:discandconcl}Discussion and Conclusions}
\cite{2000ApJ...539L.119A} presented Monte Carlo radiative transfer models for the 557 GHz $\mathrm{1_{10}}$ $\rightarrow$ $\mathrm{1_{01}}$ ground-state transition of ortho-\element[][]{H_2O} in the S140 PDR observed by SWAS. These models adopt power-law representations for their density and temperature profiles, with the mean density of \element[][]{H_2} being a factor of $\sim$ 40 larger than the density we adopted, $<$$n_\mathrm{H_2}$$>$ = 2 $\times$ $10^4$  \mbox{$\mathrm{cm}^{-3}$}, in our PDR models. They are unable to produce a single-peaked line profile with the observed integrated intensity by any combination of the ortho-\element[][]{H_2O} abundance and turbulent velocity when a static cloud is adopted. A combination of turbulent and  macroscopic motions, infall and/or outflow, is needed to reproduce the observed line profile. With our adopted density and temperature distributions, a single-peaked line profile is also not found for a static cloud. When an outflow (or infall) velocity field is incorporated, line profiles with a FWHM of $\sim$ 6 \mbox{$\mathrm{km}\ \mathrm{s}^{-1}$} are obtained for a velocity field with a gradient of a few \mbox{$\mathrm{km}\ \mathrm{s}^{-1}$}. However, when modelling the line profiles for a static embedded source, with a size $\sim$ 7 times smaller than the PDR, i.e., strongly beam-diluted in the SWAS 3.3' $\times$ 4.5' beam, we find strongly self-absorbed line profiles that remain self-absorbed for modest velocity gradients.\\
On scales of the PDR, we see that the shapes of the line profiles in the inhomogeneous case are less self-absorbed than when an homogeneous model is adopted. At smaller scales, and comparable to the size of the HIFI beam, different line profile shapes are found when zooming in on different areas of the PDR. Note that the beam of HIFI, depending on the frequency of the transition, will be $\sim$ 5 times smaller than the PDR, being able to zoom in on certain regions, and to confirm the predicted change in the line profiles. Therefore, HIFI can distinguish the embedded sources from the PDR. The former produce a completely different emission line distribution (see Table 3), with the bulk of the emission emanating in lines like, e.g., $3_{21}-2_{21}$, $3_{03}-2_{12}$, $2_{12}-1_{01}$ and $2_{20}-1_{11}$. Clearly, a combination of pure PDR and star-forming core characteristics is needed to plan for and interpret HIFI data.
\appendix
\section{\label{sec:longchameth}}
The goal of this appendix is twofold. First, we test the escape probability code \citep{2005A&A...440..559P} used in this paper. Second, we present a code based on a 'Long Characteristics Method' and compare this code with the escape probability approach.\\ 
In order to further test our escape probability code \citep{2005A&A...440..559P}, we try to reproduce the level populations and line profiles presented at the water radiative transfer workshop, held in Leiden (2004). Three test problems were presented: a static two-level problem, an expanding two-level problem, and an AGB star with 45-level ortho- and para-\element[][]{H_2O}. The first two problems have an analytical solution, see {\em http://www.mpifr-bonn.mpg.de/staff/fvandertak/H2O/radxfrtest.pdf}, and are therefore good test problems, the third is a more realistic view. Here, focus is on the first problem.\\
We consider the simple case of a two-level water molecule ($\mathrm{1_{01}\ and\ 1_{01}\ states}$) in an isothermal, constant density sphere for ortho-water abundances, relative to \element[][]{H_2}, ranging from $10^{-10}$ to $10^{-5}$. The parameters used for this model are listed in Table \ref{tab:parameters}. Level populations are calculated on a 81 $\times$ 81 $ \times$ 81 grid by means of an escape probability method as described first in \cite{2005A&A...440..559P}. Following \cite{2005A&A...440..559P}, we take the distribution of directions, $\vec{k}$, as $\vec{k}$ = $\lbrace$$-\hat{x}$,$\hat{x}$,$-\hat{y}$,$\hat{y}$,$-\hat{z}$,$\hat{z}$$\rbrace$ for a 6-way partitioned sphere (rather than $\vec{k}$ = $\lbrace$$\hat{r}$$\rbrace$) and $\beta_{ij}$(x,y,z,$\mathrm{k_i}$) = $\mathrm{{1 - e^{-\tau_{ij}(x,y,z,k_i)}}\over{6\tau_{ij}(x,y,z,k_i)}}$ for directional labels $\mathrm{k_i}$. Comparison is made with the workshop results, and our multi-zone escape probability computations are completely consistent with the analytical solutions.\\
The resulting line profiles for o-\element[][]{H_2O} abundances of $10^{-10}$, $10^{-9}$, $10^{-8}$, and $10^{-7}$ are shown in Figure \ref{fig:benchLP}. The profiles have been calculated for 1D (401 grid, slab), 2D (201 $\times$ 201 grid, cylinder) and 3D (81 $\times$ 81$ \times$ 81 grid, sphere) configurations through ray-tracing. The line profiles for the two lowest abundances, where the transition is optically thin, are single peaked, whereas at high abundances the line profile is no longer single peaked but self-absorbed, due to optical depth effects. One notices in Fig. \ref{fig:benchLP} that a 1D approximation leads to absorption effects, in case of a high abundance, that are a factor of a few larger than in the case of a 2D approximation, which in turn has bigger absorptions in comparison with a 3D case, in agreement with the results described in \cite{1997A&A...322..943J}. In case of a 2D ({\em 1D}) approximation, one forces the photons to propagate in a disc ({\em along a line}), and thus one decreases the chance for a photon to escape, leading to larger opacities, hence larger absorption effects. The 3D case is in good agreement with the analytical benchmark solutions for abundances ranging from $10^{-10}$ to $10^{-7}$.\\
\begin{table}[h!]
\begin{minipage}[b]{\columnwidth}
\renewcommand{\footnoterule}{}
\caption{Benchmark model parameters}
\label{tab:parameters}
\centering
\begin{tabular}{lc}
\hline  
Model &  \\
\hline\hline
Outer radius & 0.1 \mbox{pc}\\
Inner radius & 0.001 \mbox{pc}\\
\element[][]{H_2} density & $\mathrm{10^4}$ \mbox{$\mathrm{cm}^{-3}$}\\
Temperature gas & 40 K\\
Gaussian line profile with FWHM & 1 \mbox{$\mathrm{km}\ \mathrm{s}^{-1}$}\\
De-excitation rate coefficient & 2.18 $\times$ $10^{-10}$ \mbox{$\mathrm{cm}^{3}\ \mathrm{s}^{-1}$}\\
Excitation rate coefficient & 1.12 $\times$ $10^{-10}$ \mbox{$\mathrm{cm}^{3}\ \mathrm{s}^{-1}$}\\
No dust or velocity gradient & \\
\hline
\end{tabular}
\end{minipage}
\end{table}
In our 'Long Characteristics Method' code, photon packages travel along fixed directions on a Cartesian grid.\\
Consider an atom or molecule having {\em l} levels, with Einstein $\mathrm{A_{ij}}$, $\mathrm{B_{ij}}$ coefficients and collisional rates $\mathrm{C_{ij}}$ between levels {\em i} and {\em j}. Let ${n_{i}}$$(x,y,z)$ be the population density of the $i^\mathrm{th}$ level in point $(x,y,z)$. The equations of statistical equilibrium can then be written as:
\begin{eqnarray}
{n_{i}}(x,y,z) {\sum_{j\not= i}^l} {R_{ij}}(x,y,z) & = & {\sum_{j\not= i}^l} {n_{j}}(x,y,z){R_{ji}}(x,y,z),
\label{eq:stateq}
\end{eqnarray}
where
\begin{eqnarray}
{R_{ij}}(x,y,z) = 
\begin{cases}
{A_{ij}}\ + {B_{ij}}{\bar {J_p}}\ +\ {C_{ij}}(x,y,z) & \text{(i $>$ j)}\\
{B_{ij}}{\bar {J_p}}\  +\ {C_{ij}}(x,y,z) & \text{(i $<$ j)}
\end{cases}
\end{eqnarray}
${\bar J_p}$ is the total integrated radiation field at position p{\em (x,y,z)} in the cloud, i.e., $\bar{J}_p$ = ${\int\limits_{0}^{\infty}}\ {\bar{J}}_p^\nu \phi(\nu)d\nu$. The line profile function, $\phi_{ij}(\nu)$, determines at which frequency, i.e., position in the line profile, the line emits or absorbs, and is normalized to unity, i.e., $\int\limits_{0}^{\infty} \phi(\nu)d\nu$ = 1. When thermal as well as turbulent broadening are involved, a Gaussian line profile is produced
\begin{eqnarray}
\phi_{ij}(\nu) = \mathrm{1\over {\Delta \nu_D \sqrt{\pi}}}e^{-{\mathrm{(\nu - \nu_0)^2\over (\Delta \nu_d)^2}}},
\end{eqnarray}
with
\begin{eqnarray}
{\Delta \nu_D} = \mathrm{\nu_0 \over c}(\mathrm{2kT\over {m}} + \xi^2)^{1/2}, 
\end{eqnarray}
where $\xi$ is a root mean-square measure of the turbulent velocities. Note that the frequency shift $\Delta$$\nu$ = $\nu$ - $\nu_0$ that enters the line profile depends on the velocity field, $\Delta$$\nu$ = $\Delta$$\nu$$[$v(p),v($\mathrm{p'}$)$]$.\\
The angle integrated radiation field at frequency $\nu$ at position p{\em (x,y,z)} in the cloud is 
\begin{eqnarray}
{\bar{J}}_p^\nu & = & {\sum_{\vec{k}}}\ ^{\nu}_{\vec{k}}{I}^{tot}_p, 
\end{eqnarray}
where the sum is over all directions $\vec{k}$ in the cloud. The number of directions is arbitrary, but a 6-ray ({\em 4-ray/2-ray}) approximation is implemented for the 3D ({\em 2D/1D}) models. The total radiation field at frequency $\nu$ at position p{\em (x,y,z)} along a direction $\vec{k}$ in the cloud, i.e., $^{\nu}_{\vec{k}}{I}^{tot}_{p(x,y,z)}$, is the sum of the local contribution and the contributions coming from the other directions towards the point p{\em (x,y,z)}. In case of a 1D approximation the upward ({\em downward}) direction along $\vec {k}$ is denoted with + ({\em -}), i.e.,
\begin{eqnarray}
^{\nu}_{\vec{k}}{I}^{tot}_{p(x,y,z)} & = & ^{\nu}{I}^{loc}_{p(x,y,z)}\ +\  {\sum_{p^{'} \rightarrow +}}\ ^{\nu}_{\vec{k}}{I}^{loc}_{p^{'}}e^{-{\tau^{\nu}_{pp^{'}}}} +\  {\sum_{p^{''} \rightarrow -}}\ ^{\nu}_{\vec{k}}{I}^{loc}_{p^{''}}e^{-{\tau^{\nu}_{pp^{''}}}}
\end{eqnarray}
Here, $\tau^{\nu}_{pp^{'}}$ is the optical depth of the line at frequency $\nu$ between two points p{\em (x,y,z)} and $\mathrm{p^{'}}${\em ($x^{'}$,$y^{'}$,$z^{'}$)}. The opacity in the line, associated with the line profile, can be written as
\begin{eqnarray}
\alpha_{ij} = \mathrm{h\nu \over {4\pi}}n_j(x,y,z) B_{ji}(1 - \mathrm{n_i g_j\over {n_j g_i}})\phi_{ij}(\nu).
\end{eqnarray}
The optical depth follows by multiplying the opacity with the distance dl = p{\em (x,y,z)} - $\mathrm{p^{'}}${\em ($x^{'}$,$y^{'}$,$z^{'}$)} between the two gridpoints.
The local intensity initially carried by a photon package in a point
p{\em (x,y,z)} is given by
\begin{eqnarray}
^{\nu}{I}^{loc}_{p(x,y,z)} = \mathrm{1\over {4\pi}} A_{ij}h\nu_{ij} n_i(x,y,z)\phi_{ij}(\nu)dl + Dust + CMB,
\end{eqnarray}
with dl the size of the corresponding gridcell. \\
In this way, we use global information to determine iteratively in every gridpoint the ambient level populations. When testing this code against the benchmark model described in the beginning of this appendix we find that the level populations are completely consistent with the analytical solutions, as do the level populations calculated with our escape probability method. However, the long characteristics scheme, even though it always works, is very time consuming when a large grid is involved.\\ 
Note that the escape probability method used in this paper, as described in \cite{2005A&A...440..559P} is not a one-zone but a global approach. It has been modified such that the calculation of the level populations in a grid point depends on level populations in all other grid points through the different escape probabilities that connect any two grid points. This multi-zone implementation of the escape probability constitutes a large acceleration over long characteristics and Monte Carlo techniques.
\begin{acknowledgements}
We would like to thank Volker Ossenkopf for discussions on water observations with HIFI, and Wilfred Frieswijk for assistance with the figures. We thank the anonymous referee for his careful reading of the manuscript and his constructive remarks.
\end{acknowledgements}
\bibliographystyle{aa}
\bibliography{finalartikelII}
\end{document}